\documentclass[pra,twocolumn,floatfix,showpacs,superscriptaddress,eqsecnum]{revtex4}
\usepackage{graphicx} 
\usepackage{color}
\usepackage{amsmath}
\usepackage{amssymb}
\usepackage[urlcolor=blue, hyperindex, colorlinks, bookmarks=true,linkcolor=black,citecolor=black]{hyperref} 

\newcommand{\dg}{^\dagger}

\newcommand{\bra}[1]{\langle{#1}|}
\newcommand{\ket}[1]{|{#1}\rangle}

\newcommand{\sfrac}[2]{\mbox{$\frac{#1}{#2}$}}

\newcommand{\sch}{Schr\"odinger~}

\newcommand{\eqrf}[1]{Eq.~\eqref{#1}}

\newcommand{\ud}{\mathrm{d}}

\newcommand{\be}{\begin{equation}}
\newcommand{\ee}{\end{equation}}
\newcommand{\bea}{\begin{eqnarray}}
\newcommand{\eea}{\end{eqnarray}}

\newcommand{\mbf}[1]{\mathbf{#1}}
\newcommand{\emilycomment}[1]{\emph{\color[rgb]{.168,.549,.745}#1}}

\begin{document}

\title{Optimal control methods for fast time-varying Hamiltonians}
\date{\today}

\author{F. Motzoi}
\affiliation{Institute for Quantum Computing and Department of Physics and Astronomy,
University of Waterloo, Waterloo, Ontario, Canada N2L 3G1}

\author{J. M. Gambetta}
\affiliation{Institute for Quantum Computing and Department of Applied Mathematics,
University of Waterloo, Waterloo, Ontario, Canada N2L 3G1}

\author{S. T. Merkel}
\affiliation{Institute for Quantum Computing and Department of Physics and Astronomy,
University of Waterloo, Waterloo, Ontario, Canada N2L 3G1}

\author{F. K. Wilhelm}
\affiliation{Institute for Quantum Computing and Department of Physics and Astronomy,
University of Waterloo, Waterloo, Ontario, Canada N2L 3G1}

\begin{abstract}
In this article, we develop a numerical method to find optimal control pulses that accounts for the separation of timescales between the variation of the input control fields and the applied Hamiltonian.  In traditional numerical optimization methods, these timescales are treated as being the same. While this approximation has had much success, in applications where the input controls are filtered substantially or mixed with a fast carrier, the resulting optimized pulses have little relation to the applied physical fields. Our technique remains numerically efficient in that the dimension of our search space is only dependent on the variation of the input control fields, while our simulation of the quantum evolution is accurate on the timescale of the fast variation in the applied Hamiltonian.      
\end{abstract}


\maketitle

\section{Introduction}

Quantum information promises to significantly speed up hard computational
tasks~\cite{Nielsen00}. There is a range of candidate systems for quantum computing
hardware including atomic systems~\cite{Haeffner08}, spins~\cite{Vandersypen04}, 
and solid-state circuits~\cite{Insight}.
Quantum algorithms are typically implemented in these as a sequence
of quantum logic gates, each of which is implemented as a short temporal
pulse of an external field. These pulses have to be adapted to the
properties of the hardware at hand, which is often imperfect. This
is the task of control theory --- finding an ideal pulse shape that
implements the desired gate (or state transfer).

Control theory is rooted in applied mathematics and has been
successfully mapped to quantum physics.  The first applications of quantum control were in physical chemistry \cite{Rice00,Brumer03} and nuclear magnetic resonance (NMR)  \cite{Khaneja01, Ryan2008}, but it has since been applied to many diverse systems such as atoms \cite{Chaudhury2007}, superconducting qubits \cite{Motzoi2009,Rebentrost2009} and semiconductors \cite{hohenester06} to name a few.  There exists a wealth of analytical
and numerical methods \cite{Krotov87,Geen1991,Sklarz2002,Khaneja05,Gollub08,Nigmatullin09,Machnes2010}, the latter most often employing the GRAPE \cite{Khaneja05} or Krotov \cite{Krotov87,Sklarz2002,Nigmatullin09} algorithms.

Novel challenges arise in engineered solid-state quantum systems, such as superconducting qubits, where
quantum coherence is typically limited to microseconds, necessitating
short pulses
\cite{Lucero2008,Chow2009,Yoshihara2010,Chow2010a,Lucero2010}. Also, in these systems, the control design is limited to the
microwave frequency range where current state-of-the-art electronics restrict the variation in controls to a few nanoseconds \cite{Chow2010a,Lucero2010}.  Thus, the techniques
that are otherwise well-developed in other contexts need to be tailored to these constraints. 
Properly quantifying the shape and controllability of fast pulses
using slow modulation is an important component of lowering effective
error rates. There has been some progress in this direction with the development of optimal control methods such as DRAG \cite{Motzoi2009,Gambetta2010} and Refs. \cite{Fisher2010,Rebentrost2009}. An added benefit is that faster gates linearly speed up computation in these systems.

Most numerical control methods assume that the sampling rate of the
pulse shape that can be optimized is identical to the sampling rate
of the control field; in other words, either the control fields can be
fully shaped in real time, or we shape the envelope of a driving field represented by the Hamiltonian 
in an appropriate rotating
frame where all remnants of  time-dependence on the scale of the
driving field disappear. In both cases, the time-evolution across
a step of the shaped pulse, a pixel, can be straightforwardly approximated
using a time-independent Hamiltonian in either frame. 

In this article, we are addressing
a more general case in which the Hamiltonian across the pixel is time-dependent.  This can occur
for a wealth of reasons. If the pulse-shaping hardware has an internal filter with a bandwidth that is much smaller than the input control sample rate, then the control fields become smooth rather
than a sequence of plateaux.  Secondly, for many applications where one requires a high-fidelity operation, performing a rotating wave approximation (dropping fast rotating terms from a Hamiltonian) becomes invalid when the control pulses are extremely short.  That is, the counterrotating terms should be taken into the integration of the \sch equation even though the input controls can only be changed on a time scale which is much slower than these terms. More generally, if multiple Fourier components 
of a driving field are applied in order to, e.g., induce AC Stark-
and Zeeman-shifts, not all of these frequencies can be eliminated by a
suitable transformation. 

To account for this effect, we introduced an extra level of discretization. We separate the discretization necessary for integrating the quantum evolution from that of the discretization of the control parameters . We show with some examples that only a few extra subdivisions are needed to greatly increase the accuracy of the optimization. Moreover, since the number of controls remains the same, the search space of possible pulses does not change and the time-cost of the algorithm is only linearly affected with the number of subdivisions.

The paper is organized as follows: in
Sec.~\ref{sec:control}, we outline the control problem for optimizing quantum operations,
describe conventional approaches, and discuss their limitations; in Sec.~\ref{sec:subpixel}, we establish our coarse-grained approach to address these
limitations; specific transfer functions for piecewise cubic splines, filters, and Fourier components are given in Sec.~\ref{sec:transfer}; this methodology is applied to select problems in
section \ref{sec:applications} and concluded in Sec.~\ref{sec:conclusion}.

\section{The control problem\label{sec:control}}

In this method the standard control problem is to find a way to implement a desired unitary operation $U_{{\rm ideal}}$, given discrete controls $\mbf{u}_k$ each of length $N$. It is standard to assume that the underlying Hamiltonian is of the form
\be
\hat{H}(t) = \hat{H}_0 + \sum_k c_k(t) \hat{H}_k.\label{eq:hamil}
\ee  where $H_k$ is the control Hamiltonian and $c_k(t)$ is a continuous field which is parameterized by the controls $\mbf{u}_k$.
The goal is to choose controls $\mbf{u}_k$ such that at the final time $t=T$, the dynamics of this system approximates the target unitary operator: $U_{{\rm ideal}}$.  To find these controls one first needs to simulate the dynamics of the system, either analytically or numerically, given the Hamiltonian.  In addition, a cost function $C$ is needed to describe how close a simulated operation is to $U_{{\rm ideal}}$.  This allows us to evaluate the performance of each $\mbf{u}_k$ and thus search the space of controls for an optimum.

The most general measure for comparing two quantum
operations would be the diamond norm \cite{wat05},
\be
C = \| \mathbb{P}_Q(\mathcal{M}_{{\rm ideal}} - \mathcal{M}(T) )\mathbb{P}_Q \|_\diamond\emilycomment{,}
\ee
where $\mathcal{M}(T)$ is our simulated quantum operation and
$\mathbb{P}_Q $ is a projector onto the subspace of interest. The
ideal quantum operation is typically a unitary, formally written in Liouville space as
$\mathcal{M}_{{\rm ideal}}=U_{{\rm ideal}}\otimes\bar{U}_{{\rm ideal}}$ where $\bar{U}_{{\rm ideal}}$ is a complex conjugate.  However, in the interest of defining a more tractable search problem, we
 generally ignore the incoherent dynamics and instead require that
 the controls are time-optimal, i.e.~shorter control sequences imply more
 coherence. There are exceptions to this rule for systems with
 non-Markovian decoherence \cite{schulteherbrueggen06} or approximately decoherence-free
 subspaces \cite{Lidar1998,Zanardi1997,Nigmatullin09}. Minimizing the diamond norm
 is generally a hard problem; however, for many cases, minimizing the
 average error is sufficient. This can be done using the Frobenius-norm as the cost function, which for unitary maps is equivalent to maximizing \cite{Khaneja05,Rebentrost09}
\begin{equation}
\Phi=\frac{1}{d^{2}_Q}|\mathrm{Tr}( U_{\mathrm{ideal}}\dg U(T)\mathbb{P}_Q)|^{2},\label{eq:overlap}\end{equation}
where $d_Q$ is the dimension of the computational subspace.

The mathematical statement of the problem is to optimize $\Phi$ with respect to the vector $\mbf{u}_k$.  It is not immediately apparent how one should perform this optimization.  One of the simplest approaches is the steepest ascent (or gradient search) algorithm.  If we consider the multi-dimensional surface (control landscape) formed by $\Phi(\mbf{u}_k)$, steepest ascent is an iterative update procedure that locally examines the landscape at each iteration and provides a new $\mbf{u}_k$ by moving in the direction that increases $\Phi$ the greatest.  We take  some initial guess for the controls and iteratively update according to the rule
\be
\mbf{u}_k \rightarrow \mbf{u}_k + \mbf{\epsilon} \nabla_{\mbf{u'}_k} \Phi(\mbf{u}_k),\label{eq:update_rule}
\ee
where $\epsilon$ is a unitless step matrix (such as the inverse Hessian for BFGS-type algorthims).   Given an arbitrary initial configuration for the control fields, the algorithm follows a steepest ascent to a local optimum, at least in the case of small $\epsilon$.  For simplicity, in this work we take $\epsilon$ to be a scalar which is chosen adaptively.  It is in no way clear that this procedure will produce anything other than \emph{local} maxima, but remarkably, for the types of $\Phi$'s we examine in quantum control, the landscape is sufficiently under-constrained that gradient searches find \emph{global} maxima with high probability.  This has deep implications for the topology of the control landscape \cite{rabitz04,Fouquieres2010}.

\section{The standard algorithm}\label{sec:grape}

The GRAPE algorithm \cite{Khaneja05} is an example of a gradient search technique.  In order to numerically integrate the \sch equation it is necessary to discretize the evolution, normally into piecewise constant intervals.  The approach traditionally used in the GRAPE algorithm is to also use this discretization as the map between the continuous fields and the control vector.  That is  
\begin{equation}c_{k}(t) = \sum_{j=0}^{N-1} u_{k,j}\sqcap_j(t,\Delta t) \label{eq:digi_u},
\end{equation}
with the rectangle function
\begin{equation}\sqcap_j(t,\Delta t)\equiv\left[\Theta(t-j \Delta t) - \Theta(t-(j+1) \Delta t) \right],
\end{equation}
where $\Theta$ is the Heaviside unit step function.
Evaluating the \sch evolution using these fields and the Hamiltonian in Eq.~\eqref{eq:hamil} yields
\begin{equation}
U(0,T)  = \prod_{j=N-1}^{0}U_{j},
\label{eq:trottera}\end{equation}
where
\begin{equation}
U_{j}  =\exp\left[-i \hat{H}(j \Delta t) \Delta t\right],
\label{eq:trotterb}\end{equation}
the product running backwards is to enforce time-ordering, and $N\Delta t=T$.

 A principle insight of GRAPE  is that the gradient of $\Phi$ can be computed more efficiently than that of an arbitrary function. Specifically, the derivative of $\Phi$ with respect to one of our control variables is
\begin{equation}\label{eq:gradient}
\begin{split}\frac{\partial \Phi}{\partial u_{k,j}}  =&\frac{2}{d_Q^{\ 2}} \textrm{Re} \left[ \mathrm{Tr}\left( U_{\mathrm{ideal}}\dg\mathbb{P}_Q \frac{\partial U(T)}{\partial u_{k,j}}\mathbb{P}_Q\right)\right. \\  &\times \mathrm{Tr}\left( U_{\mathrm{ideal}}\mathbb{P}_QU(T)\dg \mathbb{P}_Q\right) \bigg],
\end{split}
\end{equation}
where
\begin{equation}
\frac{\partial U(T)}{\partial u_{k,j}}  = \left(\prod_{n=N-1}^{j+1}U_{n} \right) \frac{\partial U_{j}}{\partial u_{k,j}} \left(\prod_{n=j-1}^{0}U_{n}\right).
\end{equation}
The speedup over gradient searches based on a naive approach where one queries the fidelity function for each variation in the control values comes from the observation that the forward evolution $U(t,0)$ and the backwards evolution $U(T,t)$ need only be calculated once for a given control configuration.  This in turn allows each derivative to be calculated with a constant number of matrix multiplications, as opposed to the $N$ required for the full Schrodinger evolution, leading to an overall scaling of $O(N)$ as opposed to $O(N^2)$.

The exact form of the derivative of the unitary time slice can be found in the original GRAPE paper,
\begin{equation}
\begin{split}\frac{\partial U_{j}}{\partial u_{k,j}}  &=-i\Delta t\bar{H}_{k,j}U_{j},\\
\bar{H}_{k,j} & = \frac{1}{\Delta t}   \int_{0}^{\Delta t}e^{-i \hat{H}(j\Delta t)\tau }\hat{H}_{k}e^{i \hat{H}(j\Delta t)\tau } \ud \tau.\end{split}
\label{eq:KhanejaEq9}\end{equation}
For fine-grained control fields, i.e.~$\|\hat{H}  \Delta t\| \ll 1$, the derivative can be approximated as
\begin{equation}\frac{\partial U_{j}}{\partial u_{k,j}} =-i\Delta t \hat{H}_{k}U_{j}.\label{eq:more_grape}\end{equation}
This approximation can lead to difficulties in finding optimal solutions if the time step $\Delta t$ is not sufficiently small, but this problem can be circumvented by evaluating the integral in Eq.~\eqref{eq:KhanejaEq9} exactly \cite{Machnes2010}. Calculating this integral using the given form of $U_j$  is straightforward after obtaining the diagonalization
of $H(j \Delta t)$ (which is the preferred method of exponentiating the Hamiltonian given its hermiticity \cite{Moler03}).  Thus, the form of the derivative
of the time slice $U_j$ with respect to the $k$-th control [as in Eq.~\eqref{eq:KhanejaEq9}]  in the diagonal basis of the full Hamiltonian is

\begin{equation}
\bra{m_j}\frac{\partial U_{j}}{\partial u_{k_j}}\ket{n_j} =  \bra{m_j}  \hat{H}_{k} \ket{n_j}  \frac{e^{-i\lambda_{m_j}\Delta t} - e^{-i\lambda_{n_j}\Delta t} }{\Delta t(\lambda_{m_j} - \lambda_{n_j})}
\label{eq:exactham}
\end{equation}
for $n_j \neq m_j$, or
\begin{equation}\bra{m_j}\frac{\partial U_{j}}{\partial u_{k,j}}\ket{m_j} = -i \Delta t  \bra{m_j}  \hat{H}_{k} \ket{m_j} e^{-i\lambda_{m_j}\Delta t}\end{equation}
if $n_j=m_j$.  Here, $\lambda_{m_j}$ and $\lambda_{n_j}$ are eigenvalues of the full Hamiltonian at time slice $j$, associated with eigenvectors $\ket{m_j}$ and $\ket{n_j}$ respectively.  The actual derivative is then calculated by reverting the matrix elements back to the original, non-diagonal frame.

The gradient formula, Eq.~\eqref{eq:gradient}, explicitly assumes that the total Hamiltonian, control plus
drift, is constant across each slice.  For situations where the pulse shaping resolution limited by the pulse generator's sampling rate is slower than the Hamiltonian dynamics, it is important to consolidate the two time-scales.  Several approaches have been tried that avoid changing the existing framework.   The first is to discretize the dynamics into smaller slices and to penalize the
differences between adjacent control values.  This would lead to a search which has a dimension that scales with the smallest time-scale.  Another approach would be to coarse-grain in a way that completely discards the fast
dynamics. For example, one can ignore the counter-rotating terms in
the rotating wave approximation (RWA).  The main purpose of the rest
of the paper is to develop a method that redresses this tradeoff.

\section{The New algorithm \label{sec:subpixel}}

Our method is to separate the control parameters from the integration steps.  This, for example, occurs when the control field are combinations of Fourier components \cite{Skinner2010}, when a control field arising from an arbitrary waveform generator (AWG) is mixed with a carrier field of frequency much larger than the allowed sample rate of the AWG or when the control parameters are smoothed by a filter. In general, the field $c_{k}(t)$ can depend on all input controls $\mbf{u}_k$, though in many practical situations it only depends on local values of $\mbf{u}_k$.  Furthermore, our drift and control Hamiltonians can be time-dependent.           

\begin{figure}[t]
\centering
\includegraphics[width=0.4\textwidth]{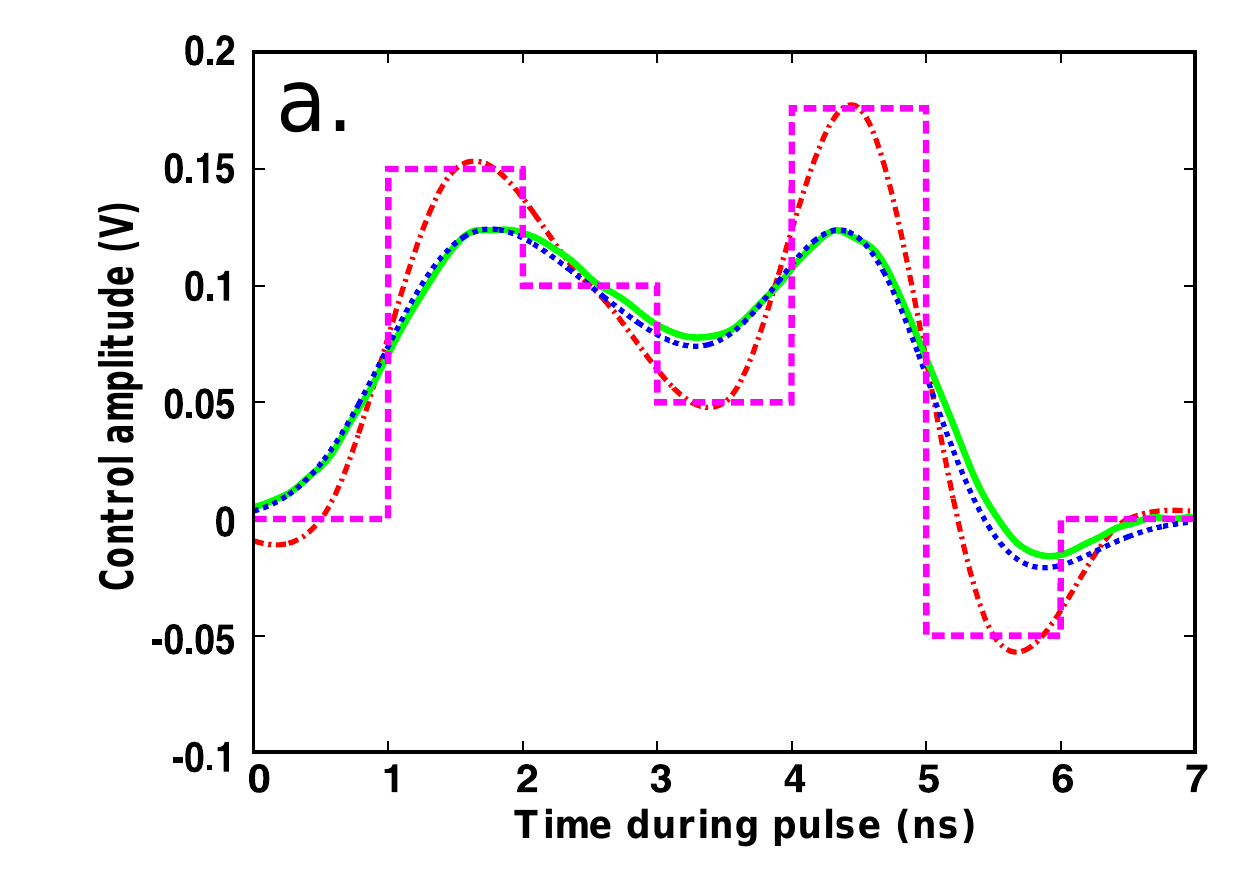}\\
\includegraphics[width=0.4\textwidth]{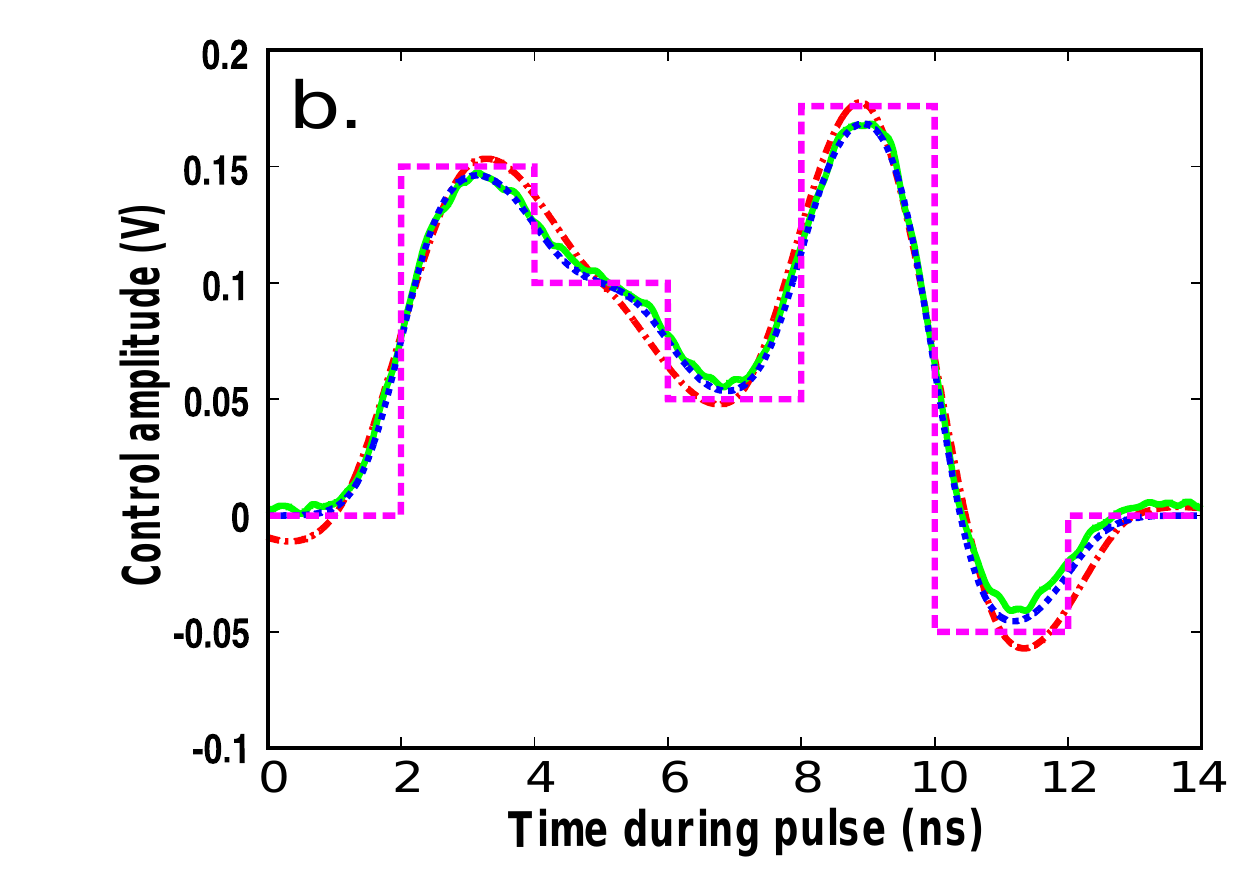}
\caption{(color online)  The purple dashed line represents the input controls $u_{j,k}$ for a arbitrary chosen 5 pixel pulse  with (a) 1ns and (b) 2ns pixels, the green solid line is the output from the Tektronix AWG5014 running at 1 GSample/s (1ns pixels) and 500 MSamples/s (2 ns pixels) respectively with 14 bit vertical resolution, the red dot-dashed line is the cubic spline interpolation, and the blue dotted line is the Gaussian filter approximation (see text for details).}
\label{envelope}
\end{figure}

As an example, we have conducted a ``theorist's experiment" to show the typical response function 
given by an AWG to a set of digital inputs. In Fig. \ref{envelope}, the dashed purple line was inputed into a Tektronix AWG5014 (the AWG) and the response (solid green line) was measured on a LeCroy WavePro 725 Zi oscilloscope. This oscilloscope had a bandwidth of 2.5 GHz (-3dB)  which was large enough to ensure that all smoothing of the pulse was from the AWG with a 250 MHz bandwidth (-3dB). The pixels were set to 1ns and 2ns in Fig. \ref{envelope} (a) and (b), respectively. Here it is abundantly clear that the input control field and the waveform produced by current state-of-the-art AWG are not the same. The remaining two lines (red dot-dashed
 and blue dotted line) represent approximations that will be discussed in Sec. \ref{sec:transfer}.

To numerically integrate the \sch equation, it is a necessity to discretize the continuous fields $c_{k}(t)$ into $s_{k,l}$, which is independented of our choice of $\mbf{u}_k$.  That is, time is digitized to $t=l \delta t$, where $\delta t$ is the time scale chosen such that $ \| \tfrac{ dH}{dt} \| \tfrac{ \delta t}{\|H\|}\ll 1$ and $M \delta t  = T$.

The evolution takes the form of Eq.~\eqref{eq:trottera} where now
\begin{equation}
U =\prod_{l=M-1}^{0}U_{l},
\label{eq:subtrotterb}\end{equation}
with the propagator,
\begin{equation}
U_{l}  =\exp\left[-i\left(H_{0,l}+\sum_ks_{k,l}H_{k,l}\right)\delta t\right].
\label{eq:subtrotterc}\end{equation}
Here, $H_{0,l}$ and $H_{k,l}$ are time-sliced versions of $H_0(t)$ and $H_k(t)$.  The update rule for $\mbf{u}_k$ is computed
the same as Eq.~\eqref{eq:update_rule}, but now the gradient of $\Phi$ is found through
\begin{equation}\frac{\partial \Phi}{\partial u_{k,j}} = \sum_{l=0}^{M-1} \frac{\partial s_{k,l}}{\partial u_{k,j}} \frac{\partial \Phi}{\partial s_{k,l}} ,\label{eq:new_update}\end{equation}
with
\begin{equation}\label{eq:subgradient}
\begin{split}\frac{\partial \Phi}{\partial s_{k,l}}  =&\frac{2}{d_Q^2} \textrm{Re} \left[ \mathrm{Tr}\left( U_{\mathrm{ideal}}\dg\mathbb{P}_Q \frac{\partial U(T)}{\partial s_{k,l}}\mathbb{P}_Q\right)\right. \\  &\times \mathrm{Tr}\left( U_{\mathrm{ideal}}\mathbb{P}_QU(T)\dg \mathbb{P}_Q\right) \bigg],
\end{split}
\end{equation}
where  
\begin{equation}
\frac{\partial U(T)}{\partial s_{k,l}}  = \left( \prod_{n=M-1}^{l+1}U_{n}\right)   \frac{\partial U_{l}}{\partial s_{k,l}}  \left( \prod_{n=l-1}^{0}U_{n}\right).
\label{eq:subgrape}\end{equation}
Calculating the $\partial U_{l} / \partial s_{k,l}$ proceeds exactly the same as in Sec.~\ref{sec:grape}, either exactly or through a linear approximation.  While this method will only ever be a linear approximation to the physical control fields, using the exact derivative as opposed to the linear approximation may speed up the gradient search.  The partial derivative $\partial s_{k,l}/ \partial u_{k,j}$ is provided by the transfer function $s_{k,l} (\mbf{u}_k)$.

\section{Transfer functions\label{sec:transfer}} 

There are a wealth of transfer functions for taking the controls
$\mbf{u}_k$ to the continuous fields $c_{k}(t)$. These transfer functions do not have to be linear but in many circumstances a linear approximation is valid. For example, the filtering seen in Fig. \ref{envelope} can is well approximated by a linear transfer function. 

A linear transfer function can be represented by  
\begin{equation}
s_{k,l} = \sum_{j=0}^{N-1} T_{k,l,j} u_{k,j},
\end{equation}
where the entirety of our transfer function is encapsulated in a transfer matrix, $T_{k,l,j}$.  In this case, evaluating Eq.~\ref{eq:new_update} is particularly straightforward since
\begin{equation}
\frac{\partial s_{k,l}}{\partial  u_{k,j} } = T_{k,l,j}. 
\end{equation} 

\begin{figure}[t]
\centering
\includegraphics[width=0.4\textwidth]{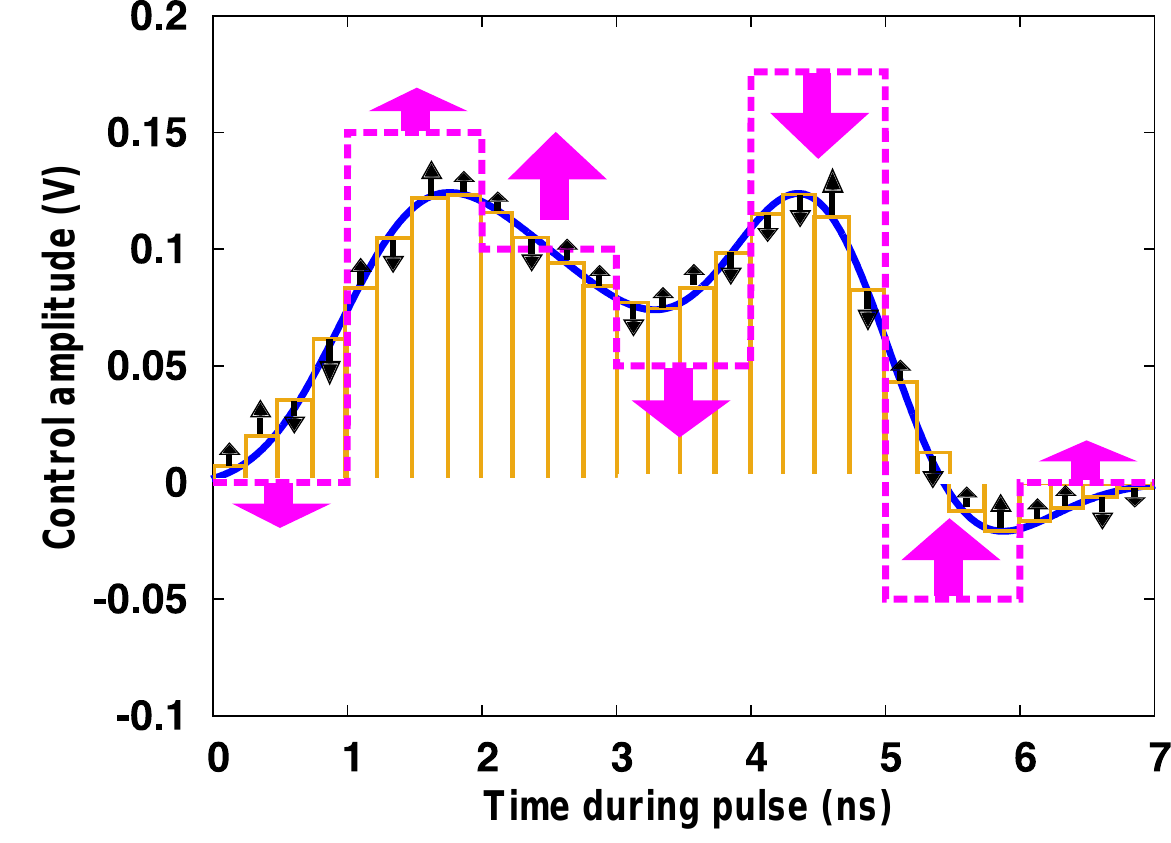}
\caption{(color online)  The purple dashed line represents the input controls pixels $u_{j,k}$, the solid blue line represents the continuous field $c_{k}(t)$, and the orange bars represent the sub-pixel approximation ($s_{k,j,l}$) to the continuous field.  The black arrows represent the update vector of the individual sub-pixels while the large purple arrows represent the update (weighted average) for the control pixels.}
\label{fig:interpol2}
\end{figure}

In the case where the input controls $\mbf{u}_k$ are indexed by time, the transfer function (such as an interpolating function, a filter, or a carrier function) will simply reparametrize the weights of the individual control pixels.  If the transfer function provides variations on a smaller time scale than the original input pixels then additional sub-pixels $s_{k,l}$ are required for the digitization of the continuous field to be valid.  This idea is illustrated in Fig. \ref{fig:interpol2}, where in particular we can think of the gradient of the (purple-dashed) control pixels as the weighted sum of the gradients (the black arrows) of the sub-pixels (orange bars) inside or near the control pixel but which we cannot directly control.  Note that this visualization works only when the control pixels are indexed in time and would not work e.g. if the controls were Fourier components.   

\subsection{Cubic spline interpolation}

Lacking any detailed description of the transfer, smoothing can be approximated by interpolating between a discrete set of points with a piecewise cubic spline interpolation.  Like the piecewise approximation in Eq.~\ref{eq:digi_u}, here we approximate the continuous fields as a piecewise continuous function
\begin{equation}
c_{k}(t) = \sum_{j=-1}^{N-1} S_{k,j}(t) \sqcap_{j+1/2}(t,\Delta t)
\end{equation}   
but now instead of remaining constant over the $\Delta t$ interval the field is described by a cubic function,
\begin{equation}
S_{k,j}(t) = \tilde{a}_{k,j} t^3 + \tilde{b}_{k,j} t^2 + \tilde{c}_{k,j} t + \tilde{d}_{k,j}.
\end{equation}
To ensure that the field and its first derivative are continuous and to introduce the control parameters, we enforce 
\begin{equation}\begin{split}
S_{k,j}((j+1/2)\Delta t)=&u_{k,j}, \\
 S'_{k,j}((j+1/2)\Delta t)=&\frac{u_{k,j+1}-u_{k,j-1}}{2 \Delta t},\\
S_{k,j}((j+3/2)\Delta t)=&u_{k,j+1}, \\
 S'_{k,j}((j+3/2)\Delta t)=&\frac{u_{k,j+2}-u_{k,j}}{2 \Delta t}.\\
\end{split}\label{label}\end{equation}
The box functions are offset by $\Delta t /2$ so that the control parameters indicate the value at the center of the step and we additionally require that the function and its derivatives are zero at the boundaries (which can be enforced by padding the control vector with zeros).

We can now derive the transfer matrix $T_{k,l,j}$ from the above conditions, and it is sparse.  For each $k,l$ pair there are only four non-zero transfer matrix elements.  There exists a $j'$ such that $(j'+1/2) \Delta t \leq l \delta t < (j'+3/2) \Delta t  $.  The non-zero matrix elements of $T_{k,l,j}$ are then
\begin{equation}
\begin{split}
T_{k,l,j'-1}=-\frac{\tau}{2 \Delta t}(\frac{\tau}{\Delta t} -1)^2\\
T_{k,l,j'}=1 + \frac{3\tau^3}{2\Delta t^3} -\frac{5\tau^2}{2\Delta t^2}\\
T_{k,l,j'+1}= \frac{\tau}{2\Delta t}+\frac{4\tau^2}{2\Delta t^2} - \frac{3\tau^3}{2\Delta t^3} \\
T_{k,l,j'+2}=  \frac{\tau^3}{2 \Delta t^3} - \frac{\tau^2}{2 \Delta t^2}
\end{split}
\end{equation}
where $\tau = l \delta t - (j'+1/2) \Delta t$.  Spline fits of this form are shown in Fig.~\ref{envelope} as the red dot-dashed line.  In these figures the spline is a better approximation for the 2ns pulse(Fig.~\ref{envelope}b) than for the 1ns pulse (Fig.~\ref{envelope}a).

\subsection{General Filters}\label{GFilter}

Most electronic systems undergo some amount of filtering.  It is important to model this to understand the shape the waveforms will take upon reaching our quantum systems.  To represent the effect of the filter the continuous field is
\begin{equation}
\begin{split}
c_k(t) =& \int_{-\infty}^{\infty} f_k(t-t')u_k(t') dt' \\
c_k(t)= &\mathcal{F}^{-1}[F_k(\omega) \mathcal{F}[u_k(t)]] ,
\end{split}\label{}\end{equation}
where $f_k(t)$, and $F_k(\omega)$ are the response functions of the
filter  for control $k$ in the time and frequency domains respectively
and $\mathcal{F}$ is the Fourier transform.  To ensure that the pulse
is zero at the boundaries we add a finite number of control parameters to
the start and end of the pulse that are fixed at zero.  The number of
parameters necessary for such a padding, $n_r$, depends on the bandwidth
of the filter, $\omega_B$, according to $n_r = \lceil 2 \pi/(\omega_B
\Delta t) \rceil$.

In the frequency domain, the transfer function is
\begin{equation}
s_{k,l}(t) = \frac{1}{2\pi}\int_{-\infty}^{\infty}u_k(t') \int_{-\infty}^{\infty}F_k(\omega)e^{i\omega(l\delta t -t')}d\omega dt'\end{equation} Using this with the piecewise constant control fields in Eq.~\eqref{eq:digi_u} and the assumption that the filter function in the frequency domain is even, the transfer matrix becomes
\begin{equation}
T_{k,l,j}=\int_{ -\infty}^{\infty } F_k(\omega)  \frac{\cos[\omega(l\delta t-(j+\sfrac{1}{2})\Delta t)]\sin[\sfrac{1}{2}\omega \Delta t]}{\pi\omega}d\omega.
\end{equation}  It is easy to precompute $T_{k,l,j}$ numerically, and for many filters it can be calculated analytically. Note also if $|l \tfrac{\delta t}{\Delta t} - j | > n_r$ we can assume that $T_{k,l,j} \approx 0$.  
     
 We demonstrate a calculation of $T_{k,l,j}$  using a Gaussian filter. In this case, the filter function is given by 
\begin{equation}
F_k(\omega)=\exp(-\omega^2/\omega_{0_k}^2) 
\end{equation} where $\omega_{0_k}$ is the reference frequency \cite{Weaver1971} for the $k$-th control. Performing the integration we find 
\begin{equation} \label{disc_gauss_convol}\begin{split}
T_{k,l,j} = \frac{1}{2} &\left\{\mathrm{erf} \left[\omega_{0_k}\left(\frac{ l \delta t-j \Delta t }{2}\right)\right]    \right. \\
&-\left.   \mathrm{erf} \left[\omega_{0_k}\left(\frac{ l \delta t - (j+1) \Delta t }{2}\right)\right] \right\} . 
\end{split}\end{equation} 

Gaussian filters are often used to approximate the actual hardware filtering typically found in experiments, which are usually parametrized by their bandwidth $\omega_B$ (frequency of 3dB attenuation). For a Gaussian the 3dB attenuation occurs at $\omega_B=0.5887\omega_0$.  Thus for the AWG with a bandwidth of $\omega_B/2\pi= 250$ MHz we find $\omega_0/2\pi = 425.4$ MHz. The Gaussian approximation to the Tektronix AWG5014 is shown in Fig.~\ref{envelope} as the blue dotted line. Here, we see that it reasonably well approximates the experiment (much better then the spline) and for the remainder of the paper, we used this as our benchmark.

\subsection{Carrier modulation}

In the above, we consider transfer functions that smooth a piecewise function into a continuous function. However, there is an interesting situation that arises when the control fields are mixed with a carrier. In this case, the transfer matrix is  
\begin{equation}\label{eq:carrier}
T_{k,l,j}= f(l\delta t) \sqcap_j(l \delta t,\Delta t),
\end{equation} where $f(t)$ is the carrier function which in most
cases takes the form of a cosine function with an arbitrary
unknown reference phase $\psi$. In appendix \ref{robust} and
Sec. \ref{carriersexample} we show how to adjust the cost function $\Phi$ to be robust against this uncertainty. Note we could also easily combine the previous smoothing transfer functions with this carrier effect.

\subsection{Fourier components}

In previous sections, we envision the control parameters as the magnitudes of piecewise constant functions, i.e.~square pulses, and the approximations as smoothed versions of these fields.  In principle, we could instead look at the $u_{k,j}$'s as more arbitrary parameters such as Fourier components (see e.g. Ref. \cite{Doria2010}), where now we write
\begin{equation} \label{cont_fourier_transform}
c_k(t) =  \sum_j u_{k,j} \sin(\omega_j t),
\end{equation}
and the transfer matrix is simply
\begin{equation} \label{inverse_fourier_transform}
T_{k,l,j} =  \sin(\omega_j  l \delta t).
\end{equation}   

\section{Applications}\label{sec:applications}

To demonstrate the usefulness of this approach, we consider three examples in this sections. The first is the implementation of a $\pi$ pulse in an anharmonic oscillator where the RWA approximation has been made. This system has been studied in detail in Ref. \cite{Steffen2003, Rebentrost09, Motzoi2009,Gambetta2010,Khani2009} and is very applicable to the superconducting qubit community. Here, we enforce that the controls have limited bandwidth (we choose a bandwidth consistent with state-of-the-art electronics) and find optimal pulses with our filtering and spline smoothing techniques. The second example is the same system but without the RWA. This allows us to demonstrate the carrier modulation technique as well as robustness of the fidelity to the initial relative phase between the carrier and the shaped envelope. The last example is a $\sqrt{\rm{iSWAP}}$ between two off-resonant qubits coupled by a $XX+YY$ interaction. This is a demonstration of multiple tones and a time-dependent drift used to couple the off-resonant qubits.

\subsection{Optimization with large filtering effects}

For the first example, we consider a system with the following Hamiltonian:
\begin{equation}\label{eq:LabHam}
\begin{split}
H=&  \left[\mathcal{E}^x(t) \cos(\omega_1 t + \psi)+ \mathcal{E}^y(t)  \sin(\omega_1 t + \psi) \right](\Gamma + \Gamma\dg)\\
 &+ \sum_{j=1,2}\omega_j \ket{j} \bra{j},
\end{split}
\end{equation} where $\omega_1$ is the $0-1$ transition frequency,  $\omega_2$ is the $0-2$ transition frequency, $\psi$ is the unknown phase of the carrier signal at the start of the pulse, $\Gamma$ is the effective lowering operator $\Gamma = \ket{0}\bra{1} +\sqrt{2} \ket{1}\bra{2}$, and $\mathcal{E}_x(t)$ and $\mathcal{E}_y(t)$ are the controls. This system represents the first three levels of an anharmonic oscillator and effectively models many systems for simulating a qubit. With this system we want to perform a $\pi$ rotation in the subspace formed by the $0$ and $1$ states without losing population to the third level. It is standard to define this $\pi$ pulse in a frame rotating at the 
frequency $\omega_1$. Moving to this frame and making the rotating wave approximation (RWA), we can model this system by the Hamiltonian 
\begin{equation}\label{eq:RWAHam}
H=  \mathcal{E}^x(t)\frac{(\Gamma + \Gamma\dg)}{2}+ \mathcal{E}^y(t)\frac{(i \Gamma - i \Gamma\dg)}{2} + \Delta\ket{2} \bra{2},
\end{equation}
where $\Delta = 2\omega_1-\omega_2$. 

\begin{figure}[t]
\centering
\includegraphics[width=0.4\textwidth]{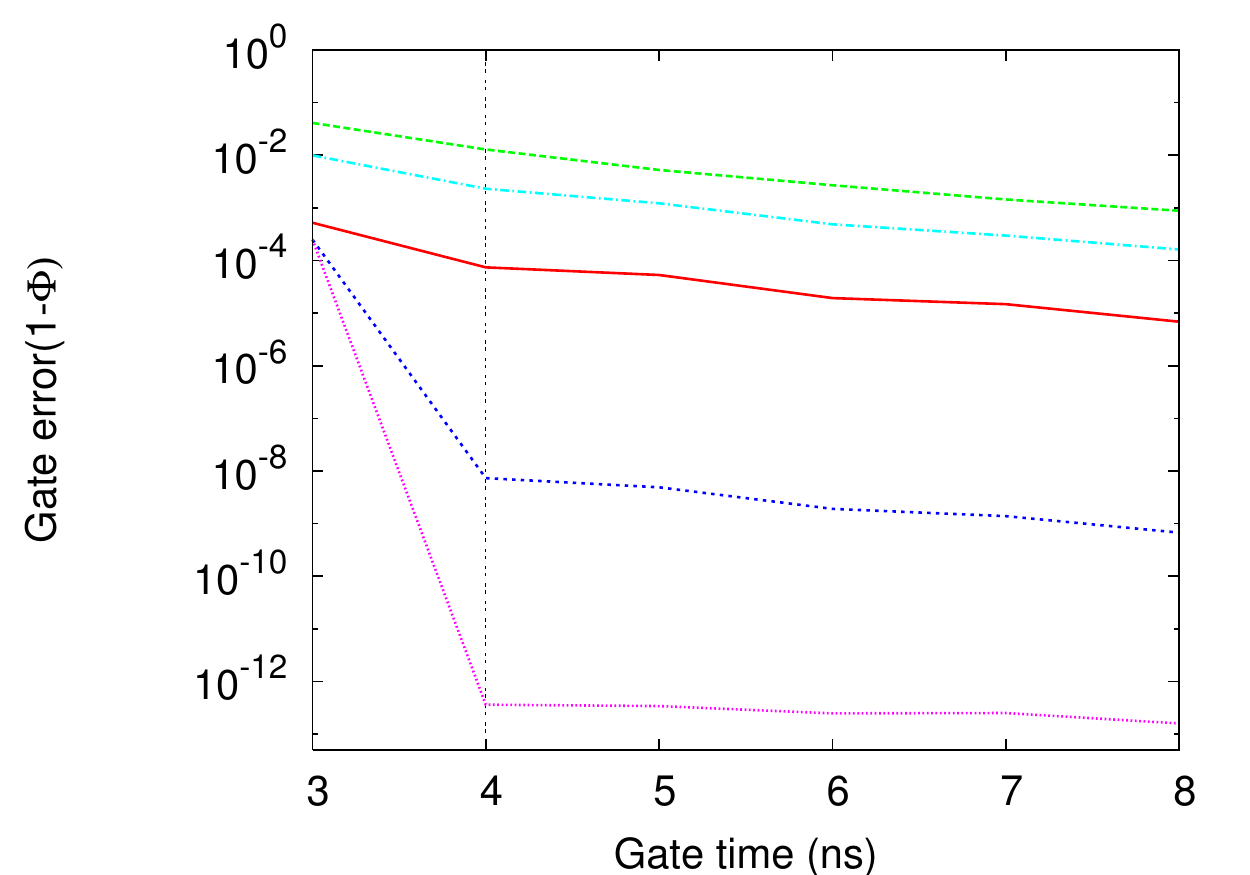}
\caption{(colour online) Gate error as a function of the gate time for the filtering example. The dotted purple line is the optimal grape solution without any filtering. The dashed green line is the predicted error when a the optimal grape solutions are filtered by a Gaussian filter approximating a AWG with a 250 MHz bandwidth. The dot-dashed cyan line is the predicted gate error after filtering with the Gaussian filter for the spline optimization with 2 sub-pixels per control pixel. The solid red line and the dotted blue line are the predicted gate errors after  filtering with the Gaussian filter for the Gaussian filter optimization with 2 and 20 sub-pixels per control pixel, respectively. The vertical dashed line indicates the gate time used in Fig. \ref{subpixels2}. Other parameters are given in the text. }
\label{subpixels}
\end{figure}

Under the RWA with two controls (with infinite bandwidth and sampling rate) it is always possible to find a solution that gives a perfect $\pi$ pulse \cite{Motzoi2009}. However, when we restrict the control parameters to 1ns (a typical setting for current microwave AWGs), we find that we need a gate time of least 4ns (4 pixels) to reach errors below $10^{-5}$. This is shown in Fig. \ref{subpixels} as the dotted purple line where we have taken $\Delta /2\pi= -500$MHz . Naively, one would then predict that these optimized pulses could be perfectly implemented. However, due to the internal filtering imposed by the AWG this is not the case (see Fig. \ref{envelope}). To demonstrate this we take the numerically optimized pulses and filter them with a Gaussian filter approximation to a AWG with an internal bandwidth of 250MHz (see Sec. \ref{GFilter}), which for short we will call the `AWG filter'. The predicted error is then shown in Fig \ref{subpixels} as the dotted green line. Here we see that the effect of the filtering is drastic, and hence the optimized pulses will not perform well for quantum operations with the AWG. 

\begin{figure}[t]
\centering
\includegraphics[width=0.4\textwidth]{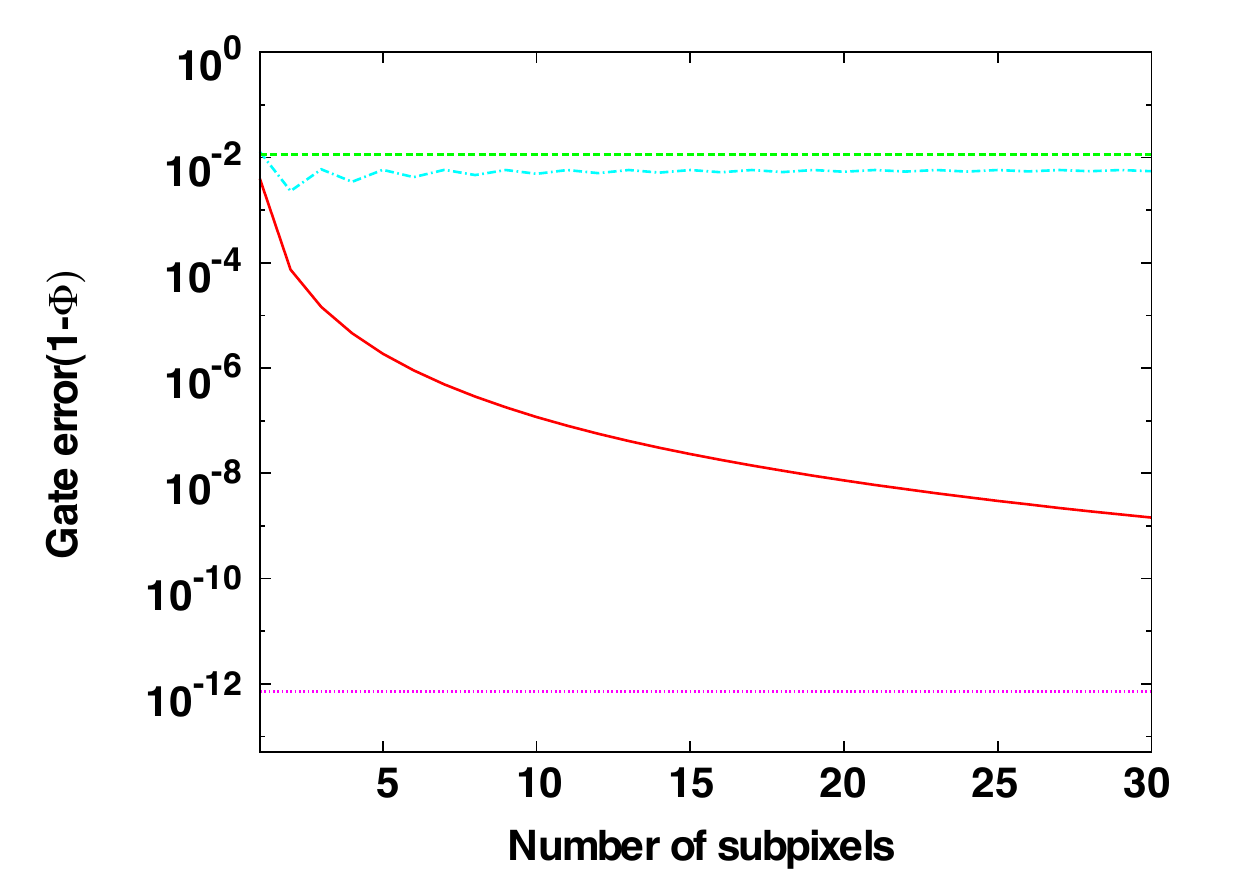}
\caption{(colour online) Gate error as a function of the number of subpixels for the filtering example. The dotted purple line is the optimal grape solution without any filtering (limited by the numerical precission). The dashed green line is the filtered optimal grape solutions when the filter is a Gaussian filter approximating a AWG with a 250 MHz bandwidth. The dot-dashed cyan and solid red lines are the predicted gate errors after filtering with the Gaussian filter for the spline and Gaussian filter optimizations, respectively. Other parameters are given in text.} 
\label{subpixels2}
\end{figure}

Taking the filtering into account during the optimization allows for much better pulses. We can do this either by finding smooth pulses with a cubic spline interpolation or by actually taking into account the filter. To demonstrate this we plot in Fig. \ref{subpixels} the error as a function of gate time when optimizing under the cubic spline interpolation (dot-dashed cyan line) and a Gaussian filter (solid red line). Here we included 2 subpixels per control pixel, and after finding the optimized solutions, applied the AWG filter. Increasing the number of subdivisions allows our algorithm to better approximate the AWG filter; this is illustrated by the dotted blue line where we find for 20 sub-pixels a greatly improved performance for all gate times. To get an indication of the performance of our algorithm with the number of subdivisions, we set the gate time to 4ns (vertical dotted line in Fig. \ref{subpixels}) and plot in Fig. \ref{subpixels2} the gate error as a function of number of sub-pixels.  Here, we observed that for only a few subdivisions the performance of our algorithm reaches very small error rates. We also find that the spline optimization is not as good as the Gaussian filter.  This is expected as we have assumed the real situation (AWG filter) to be a Gaussian filter, hence the spline optimization is not necessarily going to find pulses that are consistent with the AWG filter.  On the other hand, picking the correct transfer function improves the situation, even for very few subdivisions.

\subsection{Optimization with time-dependent carriers}\label{carriersexample}

For the second example, we take the same system and goal as the first example, but we optimize without making the RWA. Instead of focusing on the errors the arise from smoothing, we are concerned with the errors arising from the rotating terms in order to demonstrate the carrier modulation optimization. In this case, the Hamiltonian is
\begin{equation}\label{eq:RotHam}
\begin{split}
H= & \Delta\ket{2} \bra{2}+\mathcal{E}^x(t)\frac{[\Gamma (1+ e^{-2i\omega_1 t-2i\psi})+ \mathrm{h.c}]}{2}\\+& \mathcal{E}^y(t)\frac{[i\Gamma (1+ e^{-2i\omega_1 t-2i\psi})+ \mathrm{h.c}]}{2}.
\end{split}
\end{equation}
 
\begin{figure}[t]
\centering
\includegraphics[width=0.4\textwidth]{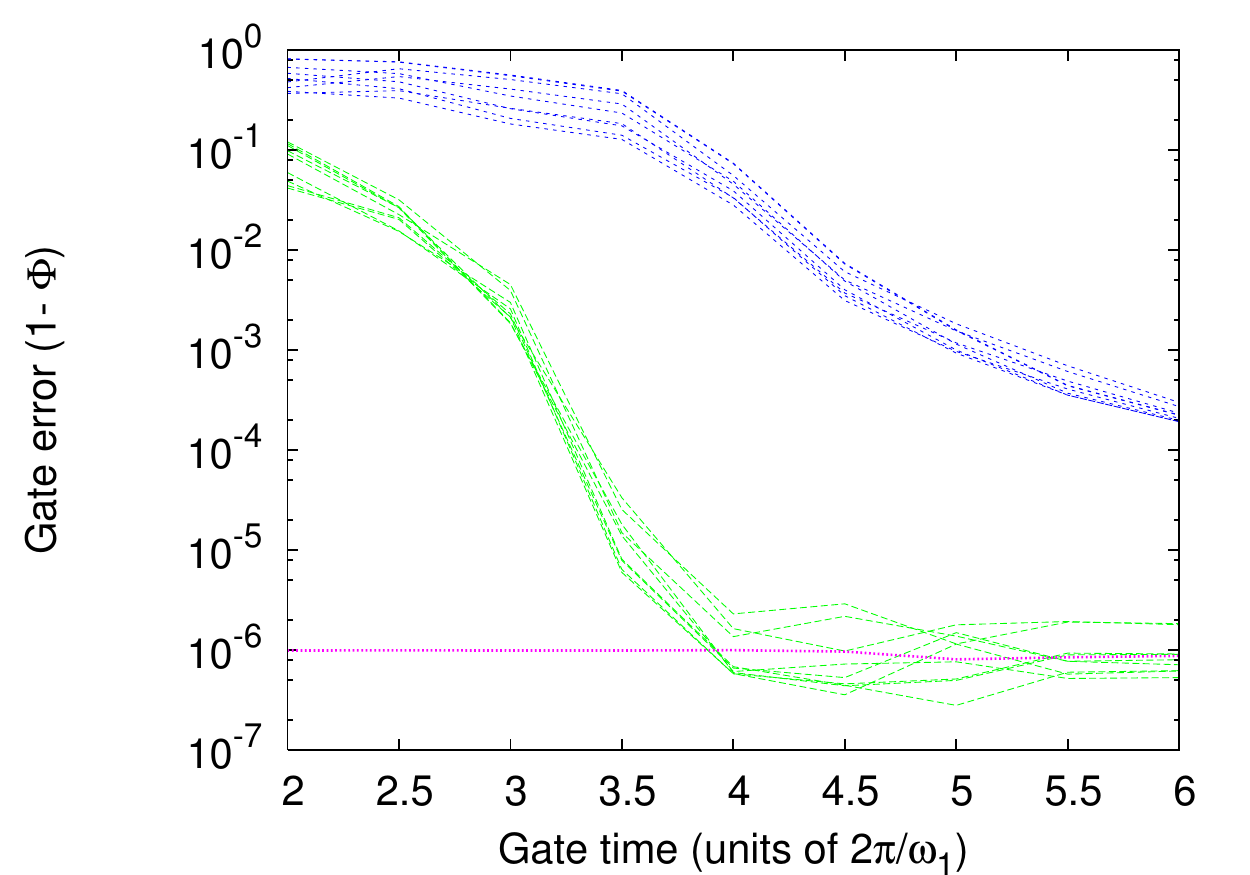}
\caption{(colour online) Gate error as a function of gate time for the carrier example. The horizontal solid purple line is the (target and achieved) gate error for the optimal solution when we make the RWA. The dotted blue lines are the predicted error when we do not make the RWA but find the optimal solution with the RWA. The lower dashed green lines are the gate errors when we optimize without making the RWA. Each line represents one of the nine different values of the relative phase between the envelope and the carrier: 2.46245, 2.13875, 1.57081, 0.304685, 0.043838, 1.65238, 0.914728, 2.02047, 0.518253. }
\label{rwainvalid}
\end{figure}

To demonstrate the error from using the wrong Hamiltonian, we first find optimal solutions for the $\pi$ pulses assuming the RWA [optimization with \eqrf{eq:RWAHam}] for $\Delta/2\pi=-500$ MHz, $\omega_1/2\pi=2.0$ GHz, and control pixels of 0.125 ns. This optimal solution is shown in Fig. \ref{rwainvalid} as the solid purple line as a function of the gate time.  For each gate time the controls are then used to evaluate what the fidelity of the operation would be if we did not make the RWA for various phases $\psi$ [evolution under Eq. \eqref{eq:RotHam}]. These results are shown in Fig. \ref{rwainvalid} as the dotted blue lines, where each line represents a  randomly chosen phase. This figure clearly indicates that for short gate times, neglecting the rotating terms leads to a large error. 

To perform the optimization with the rotating terms we simply include them as a carrier function. Since we would also like to find pulses which don't depend on $\psi$, we also use the robust technique outlined in Appendix \ref{robust}.  This is essentially an optimization of average fidelity over all possible $\psi$. In Fig. \ref{rwainvalid} we plot the gate error as a function of gate time (dashed green lines), seeing that it is possible to find pulses that robustly remove the rotation errors. Here we set the number of sub-pixels per control pixel to 100.

\subsection{Optimization with multiple tones}

The third example that we consider demonstrates that fast dynamics can occur simply from energy differences between different components in a system.  We consider two off-resonant qubits coupled by a $XX+YY$ interaction and drive with two different frequencies, one resonant with one of the qubit's transition frequency and the second tuned to exactly the average of the qubits' transition frequencies, in order to generate an $\sqrt{\rm{ISWAP}}$ operation.  Thus, we see the system has three frequencies, and there is no frame in which all fast dynamics can disappear.  Specifically, we start with the Hamiltonian of the form,
\begin{equation}\label{eq:multitonelab}
\begin{split}
H=& \mathcal{E}^{(1)}_x(t)\sigma_x^{(1)} \cos\left(\omega_d^{(1)} t \right) +  \mathcal{E}^{(1)}_y(t)\sigma_x^{(1)}\sin\left(\omega_d^{(1)} t \right)\\
&+  \mathcal{E}^{(2)}_x(t)\sigma_x^{(2)} \cos\left(\omega_d^{(2)} t \right) +  \mathcal{E}^{(2)}_y(t)\sigma_x^{(2)}\sin\left(\omega_d^{(2)} t \right)\\
&+J \left(\sigma_{+}^{(1)}\sigma_{-}^{(2)} + \sigma_{-}^{(1)}\sigma_{+}^{(2)}\right) + \omega_1\sigma_z^{(1)}+ \omega_2\sigma_z^{(1)},
\end{split}
\end{equation}
where the superscripts index the qubit, $\sigma_{\pm}$ are the raising and lowering operators, $J$ is the strength of the coupling,  $\omega_1$ and  $\omega_2$ are the qubit frequencies, and  $\omega_d^{(1)}$ and $\omega_d^{(2)}$ are the drive frequencies.  Choosing the frame rotating at the energies of the two qubits (which is conventional, but not the only available frame) and setting the first drive to $\omega_d^{(1)}=\frac{\omega_1+\omega_2}{2}$ and the second drive to $\omega_d^{(2)}=\omega_2$, we find
\begin{equation}\label{eq:multitone}
\begin{split}
H^R=& \frac{\mathcal{E}^{(1)}_x(t)}{2} (\sigma_{+}^{(1)} e^{-i\Delta t /2}+ \sigma_{-}^{(1)}e^{i\Delta t/2}) + \frac{\mathcal{E}_x^{(2)}(t)}{2}  \sigma_x^{(2)}  \\
& +  \frac{\mathcal{E}^{(1)}_y(t)}{2} (i\sigma_{+}^{(1)} e^{-i\Delta t /2}-i \sigma_{-}^{(1)}e^{i\Delta t/2})+ \frac{\mathcal{E}_y^{(2)}(t)}{2}\sigma_y^{(2)}  \\
&+J (\sigma_{+}^{(1)}\sigma_{-}^{(2)} e^{i\Delta t }+ \sigma_{-}^{(1)}\sigma_{+}^{(2)}e^{-i\Delta t}),
\end{split}
\end{equation}
where $\Delta=\omega_1-\omega_2$ is the energy difference between the qubits. 

\begin{figure}[t]
\centering
\includegraphics[width=0.4\textwidth]{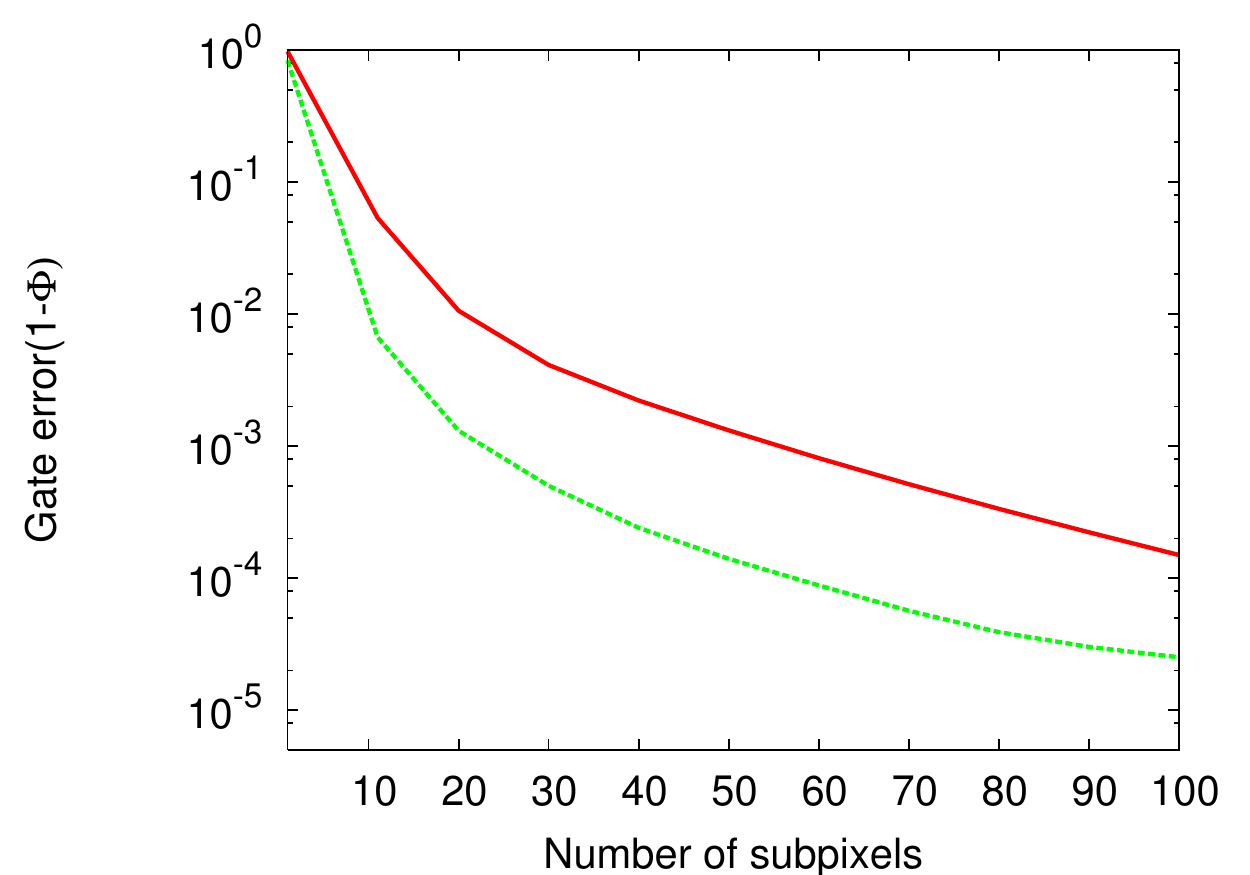}
\caption{(colour online) Gate error as a function of the number of subpixels for the multi-tone example.  The solid red line is for $\Delta/2\pi=1.0$ GHz and the green dashed line is for $\Delta/2\pi=0.5$ GHz. Other parameters are given in the text.}
\label{subpixeltimes}
\end{figure}

To illustrate the scaling with subpixels for this example we choose a gate time of $t_g=20$ns, $J/2\pi=94$ MHz, and $\Delta/2\pi= 0.5$ or $1.0$, and  in  Fig.~\ref{subpixeltimes} we plot the predicted error as a function of the number of sub-pixels for 1ns pixels.  This error is calculated by integrating the \sch equation for the optimized controls on a much finer grid. As expected, we see that for the larger detuning much more subpixels are required to achieve the same low error.

\section{Conclusions}\label{sec:conclusion}

In this article, we have developed a numerical method for finding optimal control pulses for implementing unitary gates when the control variation is at a time scale that is much different than the numerical integration time scale (set by the time variation of the Hamiltonian). This occurs in many physical examples ranging from the case when the controls are variations of an envelope function on a carrier field to the case when the controls are filtered. 
 We have shown that it is necessary to separate the input controls (occurring at the sampling rate of the waveform generator) from the fast-varying dynamics of the Hamiltonian.  We accomplish this by distinguishing the control discretization from the discretization necessary for integration. We have shown that these dynamics can be computed quickly.  Furthermore, we have given some examples of how these techniques can be applied to relevant physical systems.

\begin{acknowledgments}
We thank Adrian Lupascu and Jason Soo Hoo for boldly allowing theorists to interact with their equipment.  We acknowledge E. J. Pritchett, Th.~Schulte-Herbruggen, and B. Khani for valuable discussions. J.M.G. was supported by CIFAR, DARPA/MTO QuEST program through a grant from AFOSR, Industry Canada, MITACS, MRI and NSERC. F.M, S.T.M., and
F.K.W were supported by NSERC through the discovery grants and
QuantumWorks. This research was also funded by the Office of the Director of National
Intelligence (ODNI), Intelligence Advanced Research Projects Activity
(IARPA), through the Army Research Office.  All statements of fact, opinion
or conclusions contained herein are those of the authors and should not be
construed as representing the official views or policies of IARPA, the ODNI,
or the U.S. Government.
\end{acknowledgments}

\appendix

\section{Robust controls}\label{robust}

In many cases, we want to find controls that are robust to some uncertain parameter $\psi$ \cite{Skinner2004,Mottonen2006,Geen99}. This could be, for example, an uncontrolled phase between a carrier and the the control pixels. To account for this, we define an average performance index 
\begin{equation}\label{eq:robustphi}
\Phi=\int_\psi \Phi_\psi d\psi,
\end{equation}
where $\Phi_\psi$ is defined the same way as Eq. \eqref{eq:overlap} for each possible $\psi$. Since this is a linear combination of all the possible $\Phi_\psi$, the gradient of $\Phi$ is found by 
\begin{equation}
\frac{d\Phi}{du_{k,j}}  =\int_{\psi}\frac{d\Phi_{\psi}}{du_{k,j}}d\psi.
\end{equation} In practice the integral is replaced by summation, and we sample discretely over all possible $\psi$'s. 


\begin{thebibliography}{10}

\bibitem{Nielsen00}
M.~Nielsen and I.~Chuang, \emph{Quantum Computation and Quantum Information}
  (Cambridge University Press, Cambridge, UK, 2000).

\bibitem{Haeffner08}
H.~Haeffner, C.~Roos, and R.~Blatt, Phys. Rep. \textbf{469}, 155 (2008).

\bibitem{Vandersypen04}
L.~Vandersypen and I.~Chuang, Rev. Mod. Phys. \textbf{76}, 1037 (2004).

\bibitem{Insight}
J.~Clarke and F.~Wilhelm, Nature \textbf{453}, 1031 (2008).

\bibitem{Rice00}
S.~Rice and M.~Zhao, \emph{Optical Conrol of Molecular Dynamics} (Wiley, 2000).

\bibitem{Brumer03}
P.~Brumer and M.~Shapiro, \emph{Principles of the Quantum Control of Molecular
  Processes} (Wiley, 2003).

\bibitem{Khaneja01}
N.~Khaneja, R.~Brockett, and S.~Glaser, Phys. Rev. A \textbf{63}, 032308
  (2001).

\bibitem{Ryan2008}
C.~A. Ryan, C.~Negrevergne, M.~Laforest, E.~Knill, and R.~Laflamme, Phys. Rev.
  A \textbf{78}, 012328 (2008).

\bibitem{Chaudhury2007}
S.~Chaudhury, S.~Merkel, T.~Herr, A.~Silberfarb, I.~H. Deutsch, and P.~S.
  Jessen, Phys. Rev. Lett. \textbf{99}, 163002 (2007).

\bibitem{Motzoi2009}
F.~Motzoi, J.~M. Gambetta, P.~Rebentrost, and F.~K. Wilhelm, Phys. Rev. Lett.
  \textbf{103}, 110501 (2009).

\bibitem{Rebentrost2009}
P.~Rebentrost, I.~Serban, T.~Schulte-Herbr\"uggen, and F.~K. Wilhelm, Phys.
  Rev. Lett. \textbf{102}, 090401 (2009).

\bibitem{hohenester06}
U.~Hohenester, Physical Review B (Condensed Matter and Materials Physics)
  \textbf{74}, 161307 (2006).

\bibitem{Khaneja05}
N.~Khaneja, T.~Reiss, C.~Kehlet, T.~Schulte-Herbr\"uggen, and S.~Glaser, J.
  Magn. Reson. \textbf{172}, 296 (2005).

\bibitem{Krotov87}
V.~Krotov and I.~Feldman, Izv. Akad. Nauk SSSR, Tekh. Kinern. \textbf{2}, 160
  (1983).

\bibitem{Nigmatullin09}
R.~Nigmatullin and S.~Schirmer, New. J. Phys \textbf{11}, 105032 (2009).

\bibitem{Machnes2010}
S.~Machnes, U.~Sander, S.~J. Glaser, P.~de~Fouquieres, A.~Gruslys, S.~Schirmer,
  and T.~Schulte-Herbr\"uggen, arXiv:1011.4874v2.
  
\bibitem{Skinner2010}
T.~ E.~ Skinner and N.~ I.~ Gershenzon, J. Magn. Reson. \textbf{204}, 248 (2010).

\bibitem{Geen1991}
H.~Geen and R.~Freeman, Journal of Magnetic Resonance \textbf{93}, 93 (1991).

\bibitem{Sklarz2002}
S.~E. Sklarz and D.~J. Tannor, Physical Review A \textbf{66}, 053619 (2002).

\bibitem{Gollub08}
C.~Gollub, M.~Kowalewski, and R.~{de Vivie-Riedle}, Phys. Rev. Lett.
  \textbf{41}, 073002 (2008).

\bibitem{Lucero2008}
E.~Lucero, M.~Hofheinz, M.~Ansmann, R.~C. Bialczak, N.~Katz, M.~Neeley, A.~D.
  O'Connell, H.~Wang, A.~N. Cleland, and J.~M. Martinis, Phys. Rev. Lett.
  \textbf{100}, 247001 (2008).

\bibitem{Chow2009}
J.~M. Chow, J.~M. Gambetta, L.~Tornberg, J.~Koch, L.~S. Bishop, A.~A. Houck,
  B.~R. Johnson, L.~Frunzio, S.~M. Girvin, and R.~J. Schoelkopf, Phys. Rev.
  Lett. \textbf{102}, 090502 (2009).

\bibitem{Yoshihara2010}
F.~Yoshihara, Y.~Nakamura, and J.~S. Tsai, Phys. Rev. B \textbf{81}, 132502
  (2010).

\bibitem{Chow2010a}
J.~M. Chow, L.~DiCarlo, J.~M. Gambetta, F.~Motzoi, L.~Frunzio, S.~M. Girvin,
  and R.~J. Schoelkopf, Phys. Rev. A \textbf{82}, 040305 (2010).

\bibitem{Lucero2010}
E.~Lucero, J.~Kelly, R.~C. Bialczak, M.~Lenander, M.~Mariantoni, M.~Neeley,
  A.~D. O'Connell, D.~Sank, H.~Wang, M.~Weides, J.~Wenner, T.~Yamamoto, A.~N.
  Cleland, and J.~M. Martinis, Phys. Rev. A \textbf{82}, 042339 (2010).

\bibitem{Gambetta2010}
J.~M. Gambetta, F.~Motzoi, S.~T. Merkel, and F.~K. Wilhelm, 
Phys. Rev. A 83, 012308 (2011)

\bibitem{Fisher2010}
R.~Fisher, F.~Helmer, S.~J. Glaser, F.~Marquardt, and T.~Schulte-Herbr\"uggen,
  Phys. Rev. B \textbf{81}, 085328 (2010).

\bibitem{wat05}
J.~Watrous, Quantum Inf. Comput. \textbf{5}, 058 (2005).

\bibitem{schulteherbrueggen06}
T.~Schulte-Herbrueggen, A.~Spoerl, N.~Khaneja, and S.~J. Glaser,
  quant-ph/0609037.

\bibitem{Lidar1998}
D.~A. Lidar, I.~L. Chuang, and K.~B. Whaley, Phys. Rev. Lett. \textbf{81}, 2594
  (1998).

\bibitem{Zanardi1997}
P.~Zanardi and M.~Rasetti, Phys. Rev. Lett. \textbf{79}, 3306 (1997).

\bibitem{Rebentrost09}
P.~Rebentrost and F.~Wilhelm, Phys. Rev. B \textbf{79}, 060507(R) (2009).

\bibitem{rabitz04}
H.~A. Rabitz, M.~M. Hsieh, and C.~M. Rosenthal, Science \textbf{303}, 1998
  (2004).

\bibitem{Fouquieres2010}
P.~de~Fouquieres and S.~G. Schirmer, arxiv:1004.3492  (2010).

\bibitem{Moler03}
C.~Moler and C.~van Loan, SIAM Review \textbf{45}, 3 (2003).

\bibitem{Weaver1971}
L.~E. Weaver and D.~C. Broughto, Radio and Electronic Engineer \textbf{41}, 457
  (1971).

\bibitem{Doria2010}
P.~Doria, T.~Calarco, S.~Montangero, arXiv:1003.3750 (2010).

\bibitem{Steffen2003}
M.~Steffen, J.~M. Martinis, and I.~L. Chuang, Phys. Rev. B \textbf{68}, 224518
  (2003).

\bibitem{Khani2009}
B.~Khani , J.~M. Gambetta, F.~Motzoi, and F.~K. Wilhelm, Physica Scripta \textbf{T137}, 402
  (2009).

\bibitem{Skinner2004}
T.~E. Skinner, T.~O. Reiss, B.~Luy, N.~Khaneja, and S.~J. Glaser, Journal of
  Magnetic Resonance \textbf{167}, 68  (2004).

\bibitem{Mottonen2006}
M.~M\"ott\"onen, R.~de~Sousa, J.~Zhang, and K.~B. Whaley, Phys. Rev. A
  \textbf{73}, 022332 (2006).

\bibitem{Geen99}
H.~Geen and R.~Freeman, J. Magn. Reson. \textbf{93}, 93 (1991).

\end{thebibliography}

\end{document}